\documentclass[a4paper]{jpconf}
\usepackage{graphicx}
\usepackage{graphicx}
\usepackage{dcolumn}
\usepackage{bm}
\usepackage{bbold}
\usepackage{pslatex}
\usepackage{epsfig}
\usepackage{color}
\usepackage{dsfont}

\newcommand{\ket}[1]{\ensuremath{|#1\rangle}}

\newcommand{\be}{\begin{equation}}
\newcommand{\ee}{\end{equation}}
\newcommand{\ba}{\begin{eqnarray}}
\newcommand{\ea}{\end{eqnarray}}

\renewcommand\footnotemark{}

\begin{document}
\title{Experimental observation of weak non-Markovianity}

\author{Nadja K. Bernardes$^1$, Alvaro Cuevas$^2$, Adeline Orieux$^{2,}$\footnote[3]{Present address:
T\'el\'ecom ParisTech, CNRS LTCI, 46 rue Barrault, F-75634 Paris Cedex 13, France}, C. H. Monken$^1$, Paolo Mataloni$^2$, Fabio Sciarrino$^2$, and Marcelo F. Santos$^1$}

\address{$^1$ Departamento de F\'isica, Universidade Federal de Minas Gerais, Belo Horizonte, Caixa Postal 702, 30161-970, Brazil}
\address{$^2$Dipartimento di Fisica, Sapienza Universit\`a di Roma, Roma 00185, Italy}

\ead{nadjakb@fisica.ufmg.br}

\begin{abstract}
Non-Markovianity has recently attracted large interest due to significant advances in its characterization and its exploitation for quantum information processing. However, up to now, only non-Markovian regimes featuring environment to system backflow of information (strong non-Markovianity) have been experimentally simulated. {\color{black} In this work, using an all-optical setup we simulate and observe the so-called weak non-Markovian dynamics.} Through full process tomography, we experimentally demonstrate that the dynamics of a qubit can be non-Markovian despite an always increasing correlation between the system and its environment which, in our case, denotes no information backflow. We also show the transition from the weak to the strong regime by changing a single parameter in the environmental state, leading us to a better understanding of the fundamental features of non-Markovianity.  
\end{abstract}

\section*{Introduction}
The development of quantum technologies for information processing, communication and high resolution metrology among other applications has renewed the interest in a better understanding of the dynamics of open quantum systems. The most typical description of an open system evolution is that of a Markovian dynamics caused by the memoryless interaction of a given quantum system with its environment~\cite{breuer}. On the other hand, strong system-environment interaction, environment correlations or initial system-environment correlations may cause memory effects rendering the dynamics non-Markovian. Recently, non-Markovian dynamics has become a very trendy topic mainly due to the development of new experimental techniques for controlling and manipulating solid state and many body systems~\cite{apollaro,lorenzo,pinja1,pinja2,massimo} and also due to its possible applications in information protection and processing~\cite{verstraete,vasile,matsuzaki,chin,elsi,bogna}. 

Commonly, the evolution of a system is defined as Markovian if the corresponding quantum map is divisible in other completely positive maps (from now on CP maps), i.e. $\Lambda_{t_2,0}=\Lambda_{t_2,t_1}\Lambda_{t_1,0}$ for all $t_2\geq t_1\geq 0$~\cite{wolf, Plenio3}. For all the maps that do not satisfy this equality, the corresponding evolution is non-Markovian. The conditions for a strict Markovian dynamics are usually very hard to achieve and most experiments will present some degree of non-Markovianity. However, it is not always that the non-Markovian characteristics of a dynamics can be easily observed. Some non-Markovian processes can be identified in terms of a measurable quantity such as an increase in the distinguishability of different quantum states~\cite{breuer2} or of the entanglement between the evolving system and an ancilla~\cite{rivas}. These processes present a strong degree of non-Markovianity that has also been called essential non-Markovianity~\cite{dariusz} as opposed to a weak non-Markovianity that requires full process tomography and, therefore, is much more difficult to detect.   

It is important to understand the effects caused by such reservoirs in quantum computational systems, such as qubits, and to observe these effects in controlled experimental setups in order to pinpoint the essential mechanisms behind the different time evolutions generated by them. Hence, very recently, non-Markovianity has been investigated in different setups and contexts such as the control of the initial states of the environment~\cite{Liu, Liu2}, of its interaction with the system~\cite{tang,Steve1} or combinations of them~\cite{Fabio2}, as well as the observation of non-Markovian effects in simulated many-body physics~\cite{chiuri} or in the recovery of quantum correlations~\cite{Xu, Fabio1}. All these experiments are restricted to detecting and/or exploring the strong non-Markovianity.  

In this work we carry out an experimental characterization of the transition between weak and strong non-Markovianity. {\color{black} More specifically, the term weak non-Markovian will be used for dynamics represented by maps that are divisible in positive maps, but not in completely-positive maps. On the other hand, strong non-Markovian will be used to maps that are not even divisible in positive maps}. In particular we adopted full process tomography to observe the weak non-Markovianity dynamics of a qubit subjected to the interaction with a correlated environment. Furthermore, through the careful preparation of the environment state, the transition is induced by changing a single experimental parameter.

\section*{Theoretical model} We consider a qubit $\rho_{s}$ that interacts with an environment from which it is initially decoupled. The interaction consists of consecutive collisions each of which can produce three different effects on the system: either nothing happens, in which case the identity is applied to $\rho_{s}$ or the system is rotated by $\pi/2$ around the $X$ or $Z$ axis, undergoing a $\mathds{X}_s \equiv \sigma^s_x$ or $\mathds{Z}_s \equiv \sigma^s_z$ flip. {\color{black} The map that describes the evolution by one collision is $\Lambda_{10} (\cdot) = p_0 (\cdot) + p_x \mathds{X} (\cdot) \mathds{X} + p_z \mathds{Z} (\cdot) \mathds{Z}$ and $\{p_0,p_x,p_z\}$ are probabilities that add to one.} This is a special case of a random unitary qubit evolution. It is also a unital map.

\begin{figure*}[t!]
\centering
			\includegraphics[width=0.88\textwidth]{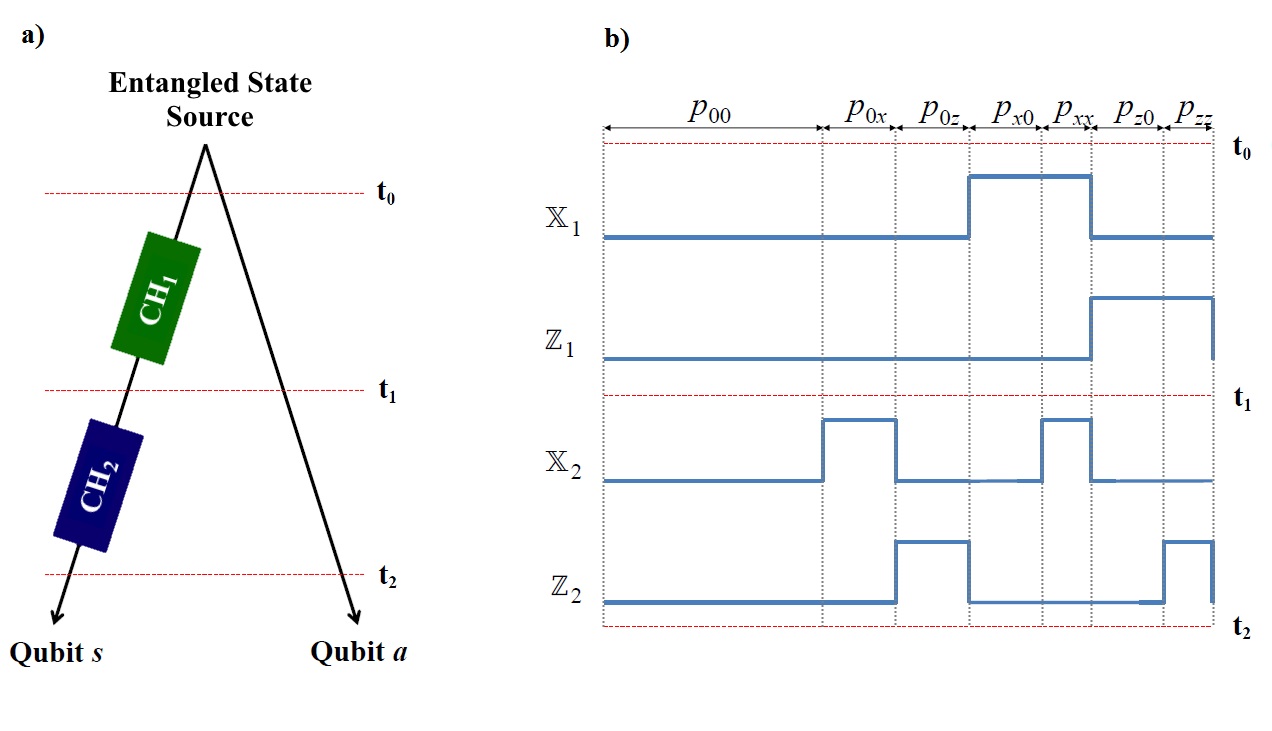} 
	\caption{\textbf{(a)} Conceptual scheme of the experiment. {\color{black} A maximally entangled state between the qubit of interest ($s$) and an ancillary state ($a$) is initially prepared at time $t_{0}$. The interaction between qubit $s$ and the environment is simulated by a sequence of two channels, performing each a  mixture of $\mathds{1}$, $\mathds{X}$ and $\mathds{Z}$ operations. The state after the interaction channel $CH_{1}$ ($CH_{2}$) is measured at $t_{1}$ ($t_{2}$). In the actual experiment the two channels are simulated each one by two liquid crystal modulators. Qubit $a$ will not suffer any change, since it is isolated from the environment. \textbf{(b)} Sequence of probabilities corresponding to  the action of the two channels, $p_{00}=(1-2\epsilon)^2$, $p_{0x}=p_{0z}=p_{z0}=p_{x0}=(1-2\epsilon)\epsilon$, and $p_{xx}=p_{zz}=2\epsilon^2$. $\mathds{X}_i$ ($\mathds{Z}_i$) is the $\mathds{X}$ ($\mathds{Z}$) operation occurring in the \textit{i}th-collision. Here we adopted the specific value of $\epsilon=0.2$.}}
	\label{scheme}
\end{figure*}

All the effects we are interested in can be observed with only two collisions described, in our case, by the general map $\Lambda_{20}(\cdot)= \sum_{mn} p_{mn}O_n O_m (\cdot) O_m O_n$ where $0\leq p_{mn}\leq 1$, $\sum_{mn} p_{mn} =1$ and $O_0=\mathds{1}$, $O_x=\mathds{X}_s$ and $O_z=\mathds{Z}_s$. Note that if the collisions are fully independent (hence non-correlated) this model is Markovian by construction and can be easily generalized for any number of collisions. The state of the system after $n$ collisions is obtained by the concatenation of single collision CP maps: $\rho_{s}(n) = \Lambda_{n0} \rho_{s}(0) = (\Lambda_{10})^n \rho_{s}(0)$. 
In this case, the joint probabilities of two consecutive flips need to respect relations such as $p_{ij} = p_{i}p_{j}$ where $i,j=\{x,z\}$.

The dynamics becomes more interesting if the collisions are correlated, i.e. when $p_{ij}\neq p_{i}p_{j}$. In particular, it is shown in Ref.~[26] that for any correlation factor $Q=\frac{p_{xx}+p_{zz}-p_{xz}-p_{zx}}{p_{xx}+p_{zz}+p_{xz}+p_{zx}}$ larger than zero, the two-collision map $\Lambda_{20}$ represents a non-Markovian evolution, i.e. $\Lambda_{20} \neq \Lambda_{21}\Lambda_{10}$ or, equivalently, $\rho_{s}(2)$ cannot be obtained by applying a CP map on $\rho_{s}(1)$. Naturally, larger values of $Q$ produce a more intense non-Markovian effect. There is, however, a transition in the type of non-Markovianity that depends on the probabilities of the flips. If correlated flips are very likely, i.e. $p_{ii}$ ($i=\{x,z\}$) is of the same order of $p_{0i}$, $p_{i0}$ and $p_{00}$, then the non-Markovianity is strong (sometimes referred to as ``essential'' in the literature~\cite{dariusz}) which means it can be witnessed by quantities such as the trace distance between different states of the system or the entanglement between the system and an ancilla. This is related to the fact that the reconstructed map $\Lambda_{21} = \Lambda_{20}\Lambda_{10}^{-1}$ is not even positive, let alone CP, i.e. it does not map the Bloch {\color{black} ball onto a set contained in it}. The extreme scenario has $p_{xx}=p_{zz}=1/2$ in which case $\rho_{s}$ after the first collision will be given by $\rho_{s}(1)=\frac{1}{2} \mathds{X}\rho_{s}(0)\mathds{X}+\frac{1}{2} \mathds{Z}\rho_{s}(0)\mathds{Z}$ and after the second collision it goes back to $\rho_{s}(0)$ (since $\mathds{X}^2=\mathds{Z}^2=\mathds{1}$).

As the flip probabilities decrease this effect becomes smaller and at some point the reconstructed map $\Lambda_{12}$ becomes positive, albeit still non-CP. For random unitary maps, as it is the case here and as it is shown in Ref.~[27], a map is divisible in positive maps if and only if the von Neumann entropy and the trace distance present a monotonic decay in time, establishing a strict relation between backflow of information in terms of these quantities and strong non-Markovianity.  {\color{black} In the case of our model, this backflow of information can also be identified observing the entanglement between system and an ancilla.} In this case, the previously discussed witnesses fail and only full process tomography of the dynamics at each step can detect the non-Markovianity character of the evolution. This situation has been defined as ``weak'' non-Markovianity in Ref.~[17]. The goal of this work is to observe this weak non-Markovianity as well as the transition of non-Markovian regimes as a function of the correlated flips probabilities. Finally, note that orthogonal flips ($\mathds{X}$ or $\mathds{Z}$) are chosen to maximize the effect but, in fact, any pair of non-commuting flips will also produce a non-Markovian map for $Q>0$~\cite{nadja}.  To implement a Kraus operator for the two collisions we exploit a mixture in time of the different Pauli operators (since the time emission of the photon is random). In our scheme we simulate non-Markovian dynamics with a classical apparatus; for more details refer to  Ref.~[25]. The conceptual scheme of the experiment we are going to show is depicted in Fig.~\ref{scheme}a). Here $CH_{1}$ and $CH_{2}$ are the interaction channels acting on the system. Fig.~\ref{scheme}b) shows the probabilities associated to the sequence of operations performed by $CH_{1}$ and $CH_{2}$.

\begin{figure}[h!]
\centering
			\includegraphics[width=0.88\textwidth]{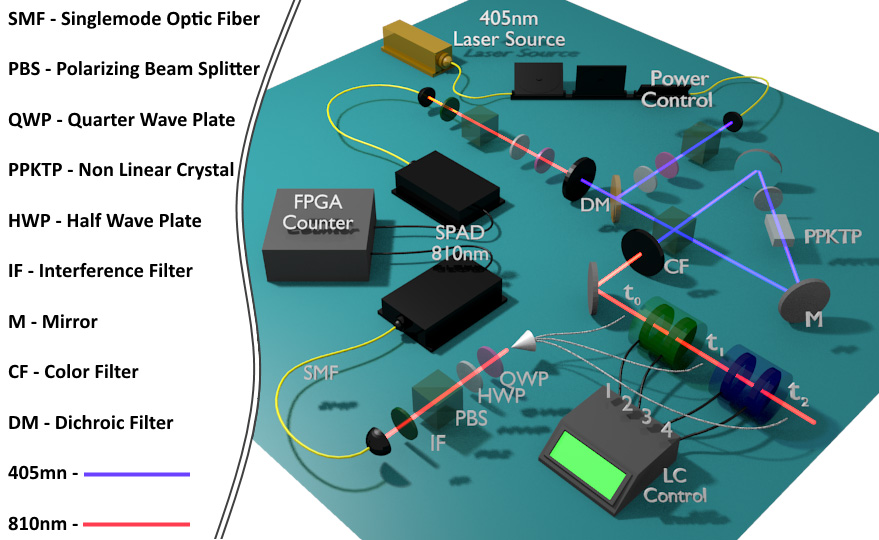} 
	\caption{{\color{black} Detailed scheme of the experiment. Twin photons are created by a polarization entanglement source. One photon (system $s$) is sent through a correlated liquid crystal environment, while the other (ancilla $a$) is let to go free. Then, the bipartite state is measured by complete state tomography at times $t_{0}$, $t_{1}$ and $t_{2}$. Liquid crystals (LCs) (two for $CH_{1}$ and two for $CH_{2}$) act as phase retarders, with the relative phase between the ordinary and extraordinary radiation components depending on the applied voltage.}}
	\label{setup}
\end{figure}

\section*{Experimental setup}
In the experiment, the system $s$ is the polarization state of an initial maximally entangled photon pair, {\color{black} $\ket{\psi_{as}}=(\ket{HV}+e^{i\alpha}\ket{VH})/\sqrt{2}$ generated by a PPKTP ultrabright source of polarization entangled photons \cite{Fedrizzi}}, where {\color{black} $\ket{H}$ ($\ket{V}$) represents the horizontal (vertical) polarization (see Fig.~\ref{setup} for details) \cite{Fedrizzi}}. Note that $\ket{i}_a\otimes\ket{j}_s$ will be simply represented as $\ket{ij}$. The environment is simulated by a sequence of four voltage {\color{black} controlled} liquid crystal  cells (LC) lying on the path of photon $s$. By a suitable control of the applied voltage on each of them, the four LCs were set to operate either as the identity or as half-wave plates. In particular the first and third LCs were oriented with the slow axis along the vertical direction, thus acting either as $\mathds{1}$ or $\mathds{Z}$, while the slow {\color{black} axis} of the second and fourth LCs were oriented along the diagonal direction ($45^{\circ}$), thus acting either as $\mathds{1}$ or $\mathds{X}$. According to the collision model, the first environment  (giving $\rho_{as}(1)$ as output result) derives from the actuation of the first two LCs (1 and 2) only, leaving the other LCs (3 and 4) in the identity regime. Finally, the second collision environment (giving $\rho_{as}(2)$ as output result) corresponds to the actuation of the four LCs in the designed way. {\color{black}  The parameter $\epsilon$ gives the probability that either $\mathds{X}$ or $\mathds{Z}$ occurred in the first collision ($p_x=\epsilon$ and $p_z=\epsilon$) and it is proportional to the time of application of the voltage to the liquid crystal. In the experimental setup it is defined by the ratio of the width of an applied voltage pulse to the width of a measurement cycle. Since the photons in our source are generated at random, the action of the liquid crystal cells driven by the voltage pulses will in fact simulate random collisions.} By controlling the time duration of the applied voltage on each LC intercepting photon $s$, it was possible to choose the probability corresponding respectively to the $\mathds{1}$, $\mathds{X}$ and $\mathds{Z}$ operations. This is given by the parameter $\epsilon$ in the following way: $p_{00}=(1-2\epsilon)^2$, $p_{0x}=p_{0z}=p_{z0}=p_{x0}=(1-2\epsilon)\epsilon$, and $p_{xx}=p_{zz}=2\epsilon^2$, as shown in Fig.~\ref{scheme}b). {\color{black} Note also that $p_{xz}=p_{zx}=0$ reflecting the fact that only perfectly correlated rotations ($Q=1$) are implemented. In this case, the theory predicts an always non-Markovian dynamics for any value of $\epsilon>0$.} In order to verify the dynamical behavior, the measurements were performed after the first collision ($t_{1}$) (in that case LCs 3 and 4 were acting as the identity and only LCs 1 and 2 were varied) and the second collision ($t_{2}$) (with all four LCs varied). As described below, in the case of strong non-Markovianity, we need just to measure an entanglement witness. However, for weak non-Markovianity, quantum state tomography should be realized.

{\color{black} In the experiment the open system dynamics is obtained by temporally mixing the three possible settings of $CH_{1}$ ($\mathds{1}$, $\mathds{X}$, or $\mathds{Z}$), giving $\rho_{as}(1)$, and by temporally mixing the seven possible settings of the action of $CH_{1}$ and  $CH_{2}$ (namely $\mathds{1}\mathds{X}$, $\mathds{X}\mathds{1}$, $\mathds{1}\mathds{Z}$, $\mathds{Z}\mathds{1}$, $\mathds{X}\mathds{X}$, $\mathds{Z}\mathds{Z}$, $\mathds{1} \mathds{1}$), giving $\rho_{as}(2)$. Note that only correlated rotations are implemented; this is done in order to maximize the non-Markovian effect as it is better explained in Ref.~[26]. The map $\Lambda_{10}$ ($\Lambda_{20}$) is obtained from the full tomographic reconstruction of $\rho_{as}(1)$ ($\rho_{as}(2)$) and the intermediate map $\Lambda_{21}$ that tells us about the character of the dynamics is calculated from $\Lambda_{21}=\Lambda_{20}\Lambda_{10}^{-1}$.}

\section*{Non-Markovian analysis}
The density matrix of a qubit state can be represented by $\rho=\left(\mathbb{1}+\vec{r}\cdot\vec{\sigma}\right)/2$, where $\vec{r}$ is the Bloch vector ($r_i=\textrm{Tr}(\rho \sigma_i)$). The action of a map $\Lambda$ on $\rho$ can be described, in general, by $\Lambda:\vec{r}\mapsto\vec{r'}=M\vec{r}+\vec{t}$, where $M$ is a matrix responsible {\color{black} for changing} the norm and {\color{black} rotating} the Bloch vector while $\vec{t}=(t_x,t_y,t_z)$ shifts its origin. For unital maps ($\vec{t}=0$), which is the case studied here, one can define {\color{black} a $4\times 4$} Hermitian matrix $\mathcal{H}=\left(\mathds{1}+\sum_{\mu,\nu=x,y,z}M_{\mu\nu}\sigma_{\mu}\otimes\sigma^{*}_{\nu}\right)/2$ so that the map $\Lambda$ is completely positive iff $\mathcal{H}\geq0$ \cite{choi,bengtsson,simon}.

\begin{figure}[t!]
\centering
			\includegraphics[width=1.0\textwidth]{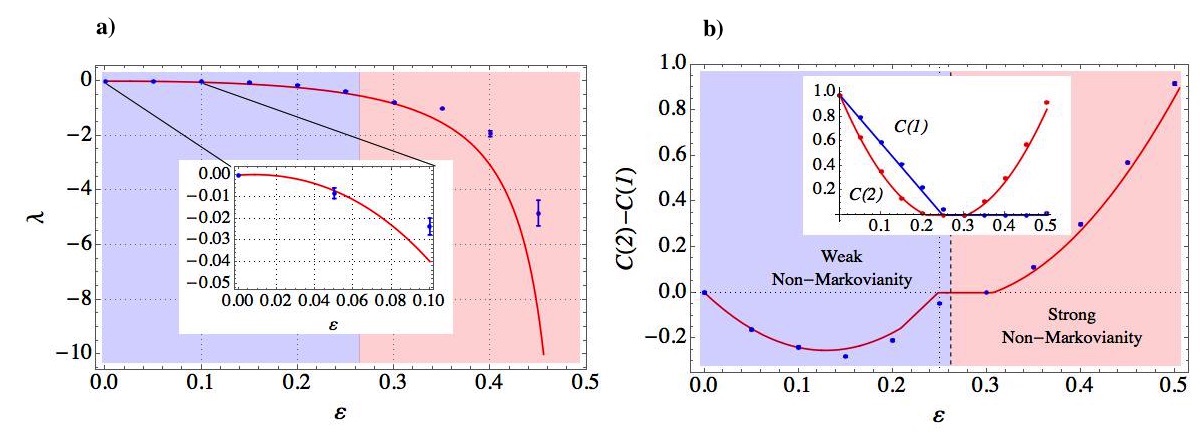} 
	\caption{\textbf{(a)} The negative eigenvalue of $\mathcal{H}$ as a function of $\epsilon$. The inset is the same curve for small values of $\epsilon$, $\epsilon<0.1$. For qualitative reasons, the weak non-Markovian regime, {\color{black}where the intermediate map is positive, but not completely positive,} is represented by the blue region and the strong non-Markovian regime{\color{black}, where the intermediate map is not even positive,} by the red region. The experimental error bars are estimated from propagation of the Poissonian statistics of photon coincidence countings in the tomographic reconstruction of the process matrix. \textbf{(b)} The difference between the concurrences of system and ancilla after two collisions $C$($2$) and one collision $C$($1$) versus $\epsilon$. The inset shows $C$($2$) and $C$($1$) versus $\epsilon$ plotted separately. For qualitative reasons, the weak non-Markovian regime is represented by the blue region and the strong non-Markovian regime by the red region. The transition from one region to the other is represented by the dashed line, which can vary its position depending on the imperfections in the experiment. The experimental error bars are estimated as explained before. The uncertainties, about $1\%$-$3\%$ of the concurrence values, are within the size of the symbols.}
	\label{figlambda}
\end{figure}

In Fig.~\ref{figlambda}a) we plot the minimum eigenvalue $\lambda$ of $\mathcal{H}$ for the map $\Lambda_{21}$ obtained from the experimental tomographic reconstruction of the system state after one and two collisions ($\rho_s(1)$ and $\rho_s(2)$ respectively). The fact that $\lambda$ is always negative for $\epsilon>0$ necessarily implies that $\Lambda_{21}$ is non-CP. As a consequence, the dynamics of the system is non-Markovian for any value in this range. The question remains whether the measured non-Markovianity is weak or strong. {\color{black} A linear map is positive iff the corresponding dynamical matrix $\mathcal{H}$ is block-positive \cite{bengtsson}. For the type of map implemented in our experiment, a simple calculation (see Methods) shows that the condition that guarantees the strong non-Markovian regime, which means that $\mathcal{H}$ is not block-positive, necessarily implies the recovery of entanglement between the system affected by the environment and an ancilla used to monitor the dynamics.} Therefore, in order to search for an eventual transition from weak to strong non-Markovian regime in our case, we have also measured the variation of the entanglement between the system $s$ and an ancilla qubit $a$ after the first and the second collision. This entanglement decreases when the system becomes more correlated with the reservoir and vice-versa, therefore, it identifies properly any backflow of information from the latter to the former. We quantify entanglement by measuring the concurrence $C$~\cite{Wooters} between system and ancilla and in Fig.~\ref{figlambda}b) we plot its difference after one and two collisions, $C(2)-C(1)$, as a function of $\epsilon$. The values of $C(1)$ and $C(2)$ are obtained from the tomographic reconstruction of the two-qubit density matrices and are ploted in the inset of Fig.~\ref{figlambda}b). Fig.~\ref{figlambda}b) shows a transition from a negative to a positive difference at around $\epsilon = 0.3$. Note that positive difference ($C(2) > C(1)$) means that system and ancilla are more entangled after two collisions than after one which, in our model, identifies strong non-Markovianity. {\color{black} As $\epsilon$ decreases, the evolution of the system will mimic a system that gets more correlated to the reservoir after the second collision (and therefore less entangled with the ancilla) which defines the regime of weak non-Markovianity} 
where $\Lambda_{21}$'s non-CP character can only be evidenced by full tomography of the map itself. Also note that the theoretical curve predicts a discontinuity in the derivative of $C(2)-C(1)$ and a plateau for a range of values of $\epsilon$. Both behaviors are easily explained if we look at the individual behaviors of $C(1,2)$ plotted in the inset of Fig.~\ref{figlambda}b). Both entanglements suddenly die~\cite{Luiz}, $C(2)$ faster than $C(1)$, but $C(1)$ remains zero for any larger value of $\epsilon$ while $C(2)$ eventually recovers due to the environmental correlations. The regions of weak and strong non-Markovianity (blue and red regions, respectively) are also presented in Fig.~\ref{figlambda}a); however, here there is no clear sign of the transition between one region to the other, so, if just the divisibility of the maps is calculated, there is no essential difference between these two types of non-Markovianity. The theoretical curves are computed assuming imperfections in the preparation of the initial state ($C(0) \sim 0.975$) and imperfect operations of the LC devices. We modelled the imperfections in the operations as $\mathds{X}_{exp}(\rho)=F \mathds{X}\rho \mathds{X}+(1-F)/2\mathds{Y}\rho \mathds{Y}+(1-F)/2\mathds{Z}\rho \mathds{Z}$ and $\mathds{Z}_{exp}(\rho)=F \mathds{Z}\rho \mathds{Z}+(1-F)/2\mathds{Y}\rho \mathds{Y}+(1-F)/2\mathds{X}\rho \mathds{X}$, and we considered $F=0.97$. 


\section*{Discussion}
We have realized experimentally the collisional model proposed in Ref.~[26] to investigate the non-Markovian dynamics of an open quantum system. We showed how the evolution of the same photonic system can transit from strong to weak non-Markovian evolution by varying only one parameter. This effect is caused by simply modulating the probability of the photon to undergo a rotation on its polarization {\color{black} state. As a result, a particular kind of} non-Markovianity which is normally not spotted in other experiments is observed here. All non-Markovianity is caused solely by {\color{black} simulating a correlated reservoir.} 
Finally, the fact that both regimes are produced by the same underlying physical mechanism explicitly shows that there is nothing necessarily fundamental about strong non-Markovian evolutions.

{\color{black} Besides its intrinsic relevance on the fundamental side, the weak non Markovianity experimentally demonstrated in this work allows to envisage future important applications regarding for instance quantum control techniques and resolution enhancement in
quantum metrology \cite{matsuzaki,chin}}.

\section*{Methods}
The positivity and completely positive character of our map can be identified by the dynamical matrix $\mathcal{H}$. For the intermediate map $\Lambda_{21}$ and error models previously explained, the reconstruction of $\mathcal{H}$ gives
\be
\mathcal{H}=\left(
\begin{array}{cccc}
h_1 & 0 & 0 & h_4\\
 0 & h_2& h_3& 0 \\
 0 & h_3 & h_2 & 0 \\
 h_4& 0 & 0 & h_1\\
\end{array}
\right),
\ee
where
\ba
h_{1}(\epsilon, F)&=& \frac{2 (F (5 F-6)+5) \epsilon ^2+3 (F-3) \epsilon +2}{2 (F-3) \epsilon +2},\nonumber\\
h_{2}(\epsilon, F)&=&-\frac{\epsilon  (2 (F (5 F-6)+5) \epsilon +F-3)}{2 (F-3) \epsilon +2}, \nonumber\\
h_3(\epsilon, F)&=&\frac{(3 F-1) \epsilon  (4 (F-1) \epsilon +1)}{2 (F-3) \epsilon +2} ,\nonumber\\
h_4(\epsilon, F)&=&\frac{\epsilon  (8 ((F-1) F+2) \epsilon +F-11)+2}{2 (F-3) \epsilon +2}. \nonumber
\ea
Its eigenvalues are
\ba
\lambda_0&=&\lambda_1=(F+1) \epsilon,\\
\lambda_2&=&-\frac{\epsilon  \left(11 F^2 \epsilon -14 F \epsilon +2 F+7 \epsilon -2\right)}{F \epsilon -3 \epsilon +1},\nonumber\\
\lambda_3&=&\frac{9 F^2\epsilon ^2-10 F \epsilon ^2+2 F \epsilon +13 \epsilon ^2-10 \epsilon +2}{F \epsilon -3 \epsilon +1}\nonumber.
\ea

First, notice that, considering that our operations are almost perfect ($F=0.97$), $\lambda_2<0$ and, as already explained, the map $\Lambda_{21}$ is not completely positive. However, we would like to identify for which values of $\epsilon$ the map becomes positive. A map is positive if the dynamical matrix is block positive, i.e. $\lambda_i+\lambda_j\geq 0$, where $\lambda_i$ are the eigenvalues of $\mathcal{H}$ \cite{bengtsson}. It is possible to show that the only inequality that does not satisfy this condition is $\lambda_0+\lambda_2$. For the case of perfect operations ($F=1$), the map will be positive only when $(1-4\epsilon)\geq 0$.

The witness that we used  to track the backflow of information was the concurrence. For our maps, it is given by
\ba
C(1)&=&\textrm{Max}[0,1-4\epsilon],\\
C(2)&=&\textrm{Max}[0,1+4 \epsilon  ((F (3 F-2)+3) \epsilon -2)].
\ea
For perfect operations ($F=1$), we can see that the condition established for positivity of the map ($(1-4\epsilon)\geq 0$) will definitely imply $C(2)-C(1)<0$. A monotonic decay of the concurrence will identify then when the map is positive, and on the other hand, non-positivity ($(1-4\epsilon)\leq 0$) implies an increase in the concurrence. However, when the operations are not perfect, there is a range of values of $\epsilon$ for which $C(2)=C(1)$, and this witness will fail to identify the exact point of transition from weak to strong non-Markovianity. Nevertheless, it will still be true that if $C(2)<C(1)$ the map is positive (weak non-Markovian) and if $C(2)>C(1)$ the map is not positive (strong non-Markovian).

\section*{Acknowledgements}
N.K.B, C.H.M. and M.F.S. would like to thank the support from the Brazilian agencies CNPq and CAPES. M.F.S. would like to thank the support of FAPEMIG, project PPM IV. This work is part of the INCT-IQ from CNPq. A.C. would like to thank the support of Comision Nacional de Investigaci\'on Cient\'ifica y Tecnol\'ogica (CONICYT). F.S. acknowledges support from CNR Programa Professor Visitante do Exterior (PVE). We acknowledge support from the European Research Council (ERC) Starting Grant 3D-QUEST (grant agreement no. 307783).

\section*{Author contributions statement}
N.K.B., C.H.Monken, and M.F.Santos proposed the original idea, F.S. and P.M. designed the experimental setup. A.C. and A.O. performed the experiment, N.K.B., A.C., and A.O. analyzed and interpreted the data. All authors contributed to the writing of the manuscript.

\section*{Additional information}

\textbf{Competing financial interests:} The authors declare no competing financial interests.
\\
\textbf{License:} This work is licensed under a Creative Commons Attribution-NonCommercial-NoDerivs 3.0 Unported License. To view a copy of this license, visit http://creativecommons.org/licenses/by-nc-nd/3.0/
\\

\end{document}